\documentclass[%
 reprint,
superscriptaddress,
 amsmath,amssymb,
prb,
]{revtex4-2}

\usepackage{graphicx}
\usepackage{epstopdf}
\DeclareGraphicsExtensions{.eps,.pdf,.png,.jpg,.gif}
\usepackage{dcolumn}
\usepackage{chemformula}
\usepackage{bm}
\usepackage{hyperref}

\begin{document}


\title{Extending Shannon’s Ionic Radii Database Using Machine Learning}
\thanks{The extended database of ionic radii is made open through \url{https://cmd-ml.github.io/}}%

\author{Ahmer A.B. Baloch}
\affiliation{Research \& Development Center, Dubai Electricity and Water Authority (DEWA), Dubai, United Arab Emirates.}%

\author{Saad M. Alqahtani}
\affiliation{Center of Research Excellence in Nanotechnology, King Fahd University of Petroleum and Minerals, Saudi Arabia}%

\author{Faisal Mumtaz}
\affiliation{Open Systems International Inc., Montreal, Quebec, Canada.}%

\author{Ali H. Muqaibel}
\affiliation{Electrical Engineering Department, King Fahd University of Petroleum and Minerals, Saudi Arabia}

\author{Sergey N. Rashkeev}
\affiliation{Department of Materials Science and Engineering, University of Maryland, College Park, MD USA.}

\author{Fahhad H. Alharbi}
\email{fahhad.alharbi@kfupm.edu.sa}
\affiliation{Center of Research Excellence in Nanotechnology, King Fahd University of Petroleum and Minerals, Saudi Arabia}%
\affiliation{Electrical Engineering Department, King Fahd University of Petroleum and Minerals, Saudi Arabia}

\date{\today}

\begin{abstract}
In computational material design, ionic radius is one of the most important physical parameters used to predict material properties. Motivated by the progress in computational materials science and material informatics, we extend the renowned Shannon’s table from 475 ions to 987 ions. Accordingly, a rigorous Machine Learning (ML) approach is employed to extend the ionic radii table using all possible combinations of Oxidation States (OS) and Coordination Numbers (CN) available in crystallographic repositories. An ionic-radius regression model for Shannon’s database is developed as a function of the period number, the valence orbital configuration, OS, CN, and Ionization Potential. In the Gaussian Process Regression (GPR) model, the reached R-square $R^2$ accuracy is 99\% while the root mean square error of radii is 0.0332 \AA. The optimized GPR model is then employed for predicting a new set of ionic radii for uncommon combinations of OS and CN extracted by harnessing crystal structures from materials project databases. The generated data is consolidated with the reputable Shannon’s data and is made available online in a database repository \url{https://cmd-ml.github.io/}.  
 
\end{abstract}


\maketitle


\section{\label{Introd} Introduction}

In computational materials design, ionic radii are essential physical features for the prediction of crystal structure and material properties \cite{R01,R02,R03}. Data-driven studies have successfully employed ionic radius to capture the physical and chemical behaviors in a variety of applications including crystallographic nature \cite{R04,R05,R06}, batteries \cite{R07,R08}, scintillators \cite{R09}, semiconductor absorbers \cite{R10,R11,R12}, seawater properties \cite{R13} and mineralogy \cite{R14,R15}. In this background, computing the ionic radius for arbitrary oxidation states and coordination geometries to study material properties is of a considerable theoretical and applied interest. 

Ionic radius is not a fixed value for a particular ion but it changes with oxidation state, coordination environment, and orbital configurations among other properties. It is defined by the distance between the nucleus of a cation (anion) and its adjacent anion (cation) in a crystal structure. However, calculating the ionic radii is a complicated problem as the electron distribution is probabilistic and forms clouds without clear boundaries to demarcate different ions. Therefore, many scientists have worked on reasonable ways of defining the ionic radii of different chemical elements \cite{R16,R17,R18}. The most influential work was carried out by Shannon who predicted and compiled the ionic radii data for common oxidation states (OS) and coordination numbers (CN) \cite{R16}. It resulted in the most acceptable databases of ionic radii collection from Refs. \cite{R16,R17,R18}. Albeit this data has been used extensively in different fields, the Shannon’s table is incomplete, as it does not cover the entire periodic table due to a lack of structural information for all common and uncommon OS with all possible coordination CN. Moreover, the absence of this data has rendered difficulties in predicting and screening novel compounds. For instance, most of the material searching for new perovskite absorbers is restricted only for cations and anions available in Shannon’s ionic radii collection \cite{R11,R19,R20,R21,R22,R23}. Researchers designing halide perovskites have been using a wide range of effective ionic radius values for tin ion \ch{(Sn^2^+)}  (from 0.93 \AA $\,$ to 1.36 \AA) \cite{R20,R21,R22,R23} resulting in misleading analysis in numerous case studies. Therefore, extending Shannon’s database could provide missing ion information for different materials. Fortunately, during the last decades, computational material science has evolved rapidly thanks to useful crystal prediction tools such as Density Functional Theory (DFT) and expanding databases for crystal structures \cite{R24}. Moreover, a huge amount of experimental data has been accumulated that can be employed for empirical correlations and computational materials design \cite{R25}. The ideal design or model should be able to connect any type of physical and chemical properties of a compound to its constituent parameters \cite{R26}. Driven by the growth of material informatics, we developed a robust ionic radii model for a complete periodic table and tabulated the data of the missing ions in the already existing databases for the research community \cite{R27,R28,R29,R30}. The predicted Shannon’s ionic radii can be used for classifying crystal structures, tolerance factors, geometrical properties, etc. This research is timely needed and relevant to the evolving material informatics field; but it will also find applications in many other areas. 

Theoretically, the ionic radius is a fundamental property of the atom that gains or loses an electron from its valence shell. The contribution from the valence electron modifies ionic radius in two ways: (1) repulsions between valence electrons are changed due to the increase/decrease in the number of electrons, and (2) the effective nuclear charge experienced by the remaining core electrons is altered by adding or removing valence electrons. The seminal related works can be traced back to Goldschmidt \cite{R31}, followed by Pauling \cite{R18} and Zachariasen \cite{R32}. Wasastjerna \cite{R33} originally calculated the radius of ions using their relative volumes as measured from optical spectroscopy. Pauling introduced an effective nuclear charge to consider the distance between ions as a sum of an anionic and a cationic radius with a fixed radius of 1.40 \AA $\,$ for \ch{O^2-} ion \cite{R18}. A comprehensive analysis of crystal structures then subsequently led to the publication of the updated ionic radii by Shannon \cite{R16}, with specific CN and different OS. To be consistent with Pauling's ionic radius values, Shannon had used an ionic radius of 1.40 \AA $\,$ for \ch{O^2-} and called it an “effective” ionic radius. The effective ionic radii for nitrides was then calculated by Baur \cite{R34} and for sulfides/ fluorides by Shannon \cite{R35,R36}. Zachariasen \cite{R37} developed a functional form for calculating the bond length for oxygen and halogen compounds of $d$-orbital and $f$-orbital elements. The effective ionic radii of the trivalent and divalent rare-earth ions were predicted by Jia \cite{R38} who calculated unknown radii for different coordination numbers, ranging from 6 to 12, for $4f$ orbital elements. Recently, the effective ionic radii for anions have been calculated for binary alkali compounds via accessing a subset of suitable crystals from materials project \cite{R39}. Ouyang carried out a comprehensive study for designing perovskite materials using ionic radii based ML descriptor \cite{R40}.

Shannon derived the empirical ionic radius, called effective ionic radii, by systematically reproducing mean experimental cation-anion distances in crystal structures using the equation, $R_{(i,anion)}+R_{(i,cation)}=d_{(anion-cation)}$ where $R_{(i,)}$  is the ionic radii and $d$ is the interatomic distance. The data derived by Shannon was formulated for 1000 average interatomic distances and empirical bond-length/bond-strength values \cite{R35}. Corrections to the radii were carried out for physical parameters using correlations between: (1) ionic radii and unit cell volume; (2) ionic radii and CN; (3) ionic radii and OS, and; (4) ionic radii and orbital configuration \cite{R16}. However, the main limitations in Shannon’s data arise from its origin in using primarily the oxide ion and hence poses a challenge when computing cation radii for other anions. To address the issue of different anions, one can use the concept of difference in empirical ionic radii \cite{R41} as different approaches for calculating ionic radius give relative values with similar trend due to geometric nature of ions. To make this comparison, one can subtract the oxide radius (\ch{O^2-}) \cite{R42} from another anion radius (say sulfide ion, \ch{S^2-}). Negative difference would then suggest that sulfide cation-anion distances are smaller than in the oxide of the same element \cite{R36,R41}.  

Rationally, it is noticed from the literature that ionic radii depend on many features, and their calculation requires to take into account corrections for OS and CN. Furthermore, for better comprehension of the ionic radius and its relationship to the physical and chemical properties, interaction between ions and their electronic shells should be accurately taken into account using valence electrons in $s, p, d, f$ orbital configurations. In addition, most of the literature work involves interpolation or extrapolation without making a generalized model based on physical descriptors \cite{R34,R37,R38,R39,R43}. Nonetheless, it is noticed that the differences between various methods are not random and follow particular trends. Therefore, for materials informatics, it is important to use a single standard. 

\begin{figure*} [htp]
\includegraphics[width=0.9\textwidth]{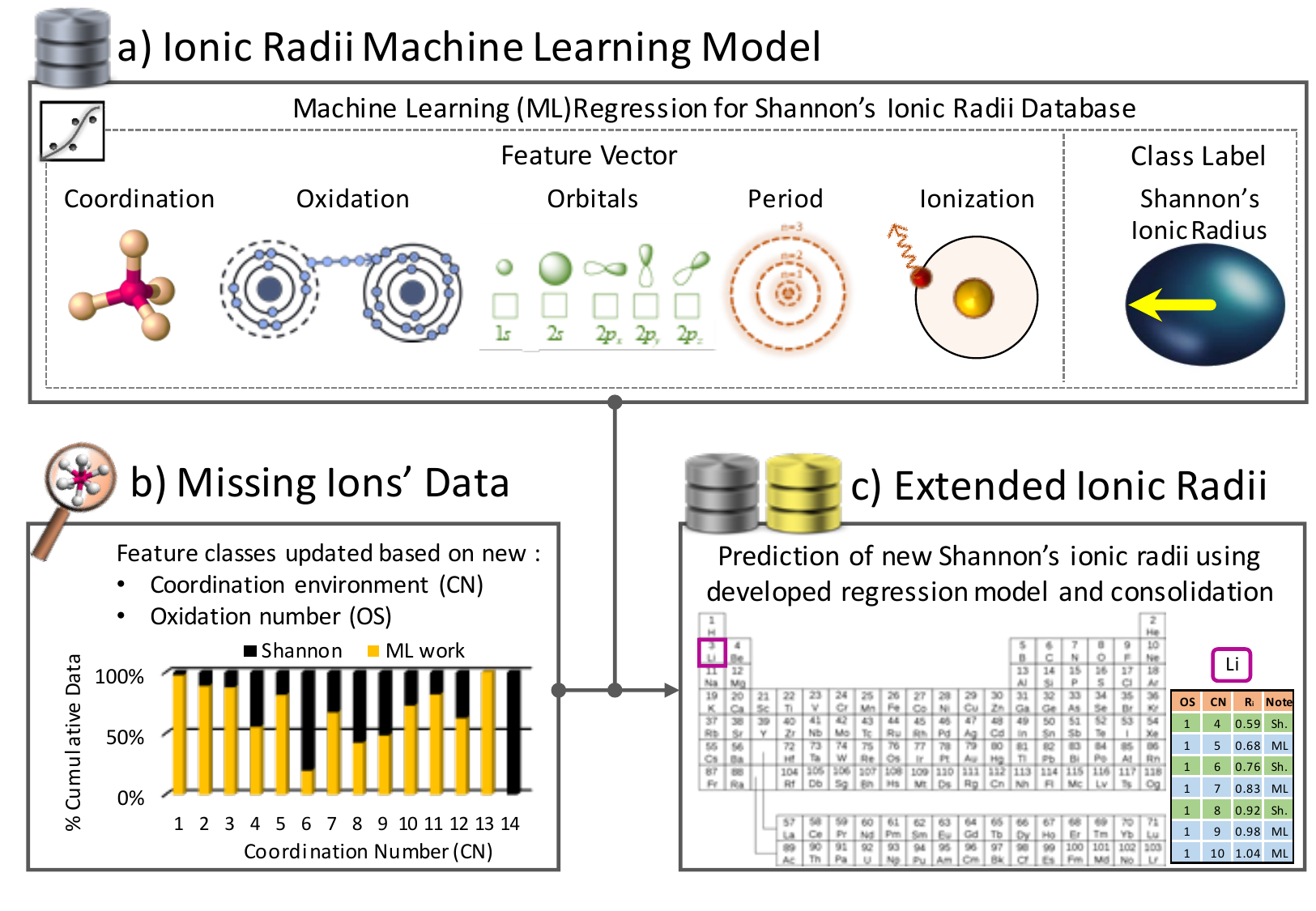}
\caption{Workflow for extending the Shannon's ionic radii database using machine learning and data harvesting. a) Ionic radii machine learning model for existing Shannon’s database using physically-guided features. b) Collection of missing ions data from material repositories and its comparison with Shannon’s database in terms of Coordination Number, CN: 1--14, and c) Consolidation of both Shannon table and missing species’ ionic radii databases from predicted values using missing ions’ data and regression model.}
\label{Fig1}
\end{figure*}

Accordingly, we present a rigorous Machine Learning (ML) approach to extend the ionic radii table of Shannon’s database using all possible combinations of OS and CN available in material informatics repositories \cite{R27,R30}. Rare oxidation states and coordination environments, as well as those missing in Shannon’s database, were considered. Data for different combinations of OS and CN were carefully harvested from the crystallographic database for 7969 crystal structures using material repositories with the original data stemming from an experimental databases \cite{R27,R28,R44}. Fig. \ref{Fig1} shows the flowchart adopted for the regression problem with selected features. Regression on the Shannon's ionic radius as a function of the period number in the Periodic Table of Elements, OS, CN, Ionization Potential ($E_{IP}$) and the valence orbital configuration ($s, p, d, f$) was performed. The developed model is valid for all elemental families and not just limited to specific classes. The descriptors presented here are relatively simple and physically intuitive, relying only on eight fundamental parameters to describe an ion of any element in a periodic table. Several state-of-the-art ML-based regression models were employed including Linear Regression (LR), Support Vector Machines (SVM), Decision Trees (DT), and Gaussian Process Regression (GPR). All the used methods worked satisfactory; however, the GPR model showed the best predictive accuracy. For training and testing, seven-fold cross-validation was performed for the Shannon’s table \cite{R16}. By optimizing the hyperparameters of the ML algorithms, Gaussian Process Regression (GPR) showed the minimum Root Mean Square Error (RMSE) of 0.0332 \AA with a  Coefficient of Determination ($R^2$) reaching 99.3\%. These results illustrate effective implementation of the regression model which was then employed for predicting a new set of ionic radii for uncommon combinations of OS and CN. We extended the Shannon’s ionic radii table from 475 to 987 ions by predicting ionic radii for 512 new compounds. The generated data was then consolidated with the reputable Shannon’s table as shown in Fig. \ref{Fig2}. The newly developed table should assist accurate prediction of crystal structures by considering the ionic radius value based on the exact OS/CN, rather than the common OS/CN, which translates to better prediction of material properties. The resulting data has been made available online in open database repositories for research.

\section{\label{Methodology} Methodology}
Prediction of Shannon's ionic radii (denoted as $R_i$ in this paper) , for missing materials and ions in Shannon’s database was performed using supervised ML and data harvesting as described in Fig. \ref{Fig1}. The regression model for all periodic table elements was evaluated on a test/train split approach with 7-fold cross-validation. Statistical correlation for strength and direction of the feature/target vector was evaluated using Spearman’s rank coefficient ($\rho$). For data mining, missing ions’ data in terms of coordination numbers and oxidation states were carefully collected from the ICSD, initially rooting from the experimental data \cite{R27}. The developed regression model is applied for the prediction of new ionic radii based on the recently harvested coordination environments and oxidation states from Materials Project. The predicted data is then consolidated with existing databases to extend the ionic information in tabulated form for ease of use for the research community.

\subsection{Data and Features}
Shannon’s dataset contains about 475 ions \cite{R16}. The table includes the ion, oxidation (formal charge), coordination and ionic radius. We selected physically-guided features for establishing a regression model for the target function of Shannon’s ionic radius as shown in Fig. \ref{Fig1}a. The features based on the nature of the ionic radius are listed below:
\begin{itemize}
\item Atomic properties
    \begin{itemize}
        \item 	Period number,
	    \item   $s$-orbital outer shell valence electrons,
	    \item   $p$-orbital outer shell valence electrons,
	    \item   $d$-orbital outer shell valence electrons,
	    \item   $f$-orbital outer shell valence electrons,
	    \item   $e^{-E_{IP}}$, Ionization Potential (negative exponent).
    \end{itemize}
\item Ionic properties
    \begin{itemize}
        \item   Oxidation State (OS),
        \item   Coordination Number (CN).
    \end{itemize}
\end{itemize}

Data for elemental properties were extracted from the web of elements \cite{R45} whereas OS, CN and $R_i$ were adopted from the Shannon’s compilation \cite{R16} which includes Pauling \cite{R18} and Ahrens \cite{R17} as well. The descriptors presented here are relatively simple, having only eight parameters to describe an ion of any element in the periodic table. However, it has not been benchmarked for high or low spin materials as the number of counts in the original Shannon’s table is low to make any statistical significance for such a model. Orbital outer shell valence electrons as a feature can be explained by considering an example of Scandium with orbital configuration, \ch{Sc}:$1s^2$ $2s^2$ $2p^6$ $3s^2$ $3p^6$ $4s^2$ $3d^1$. When condensed [\ch{Sc}] using noble gas configuration it becomes [\ch{Ar}] $4s^2$ $3d^1$ and accordingly our  input feature vector for [$s$ $p$ $d$ $f$] would be [2 0 1 0].

For a detailed clarification of the materials’ data acquisition methodology and harnessing data for 7969 oxides, the Materials Project Representational State Transfer (REST) Application Programming Interface (API) was interfaced using Python Materials Genomics (pymatgen) library \cite{R30,R46}. The raw computed data was acquired based on user-defined criteria and then utilized to perform post-processing analysis to derive further properties of the materials via pymatgen library. The pymatgen is an open-source library that has several packages such as core, electronic\_structure, entries, io, etc. In this work, initially, materials space of the considered 7982 oxides in the work of Waroquiers et. al. \cite{R49} were used as an input criterion to obtain the matching existed data in the Materials Project database \cite{R30} using query class, which is based on MongoDB-like syntax. After verifying their existence in ICSD as either experimental or theoretical data, we were left with 7969 oxides. As CN is a crucial physical-attribute in our supervised-predictive model, the corresponding CNs were acquired from the work of Waroquiers et. al. \cite{R49}.

\begin{figure*} [htp]
\includegraphics[width=0.9\textwidth]{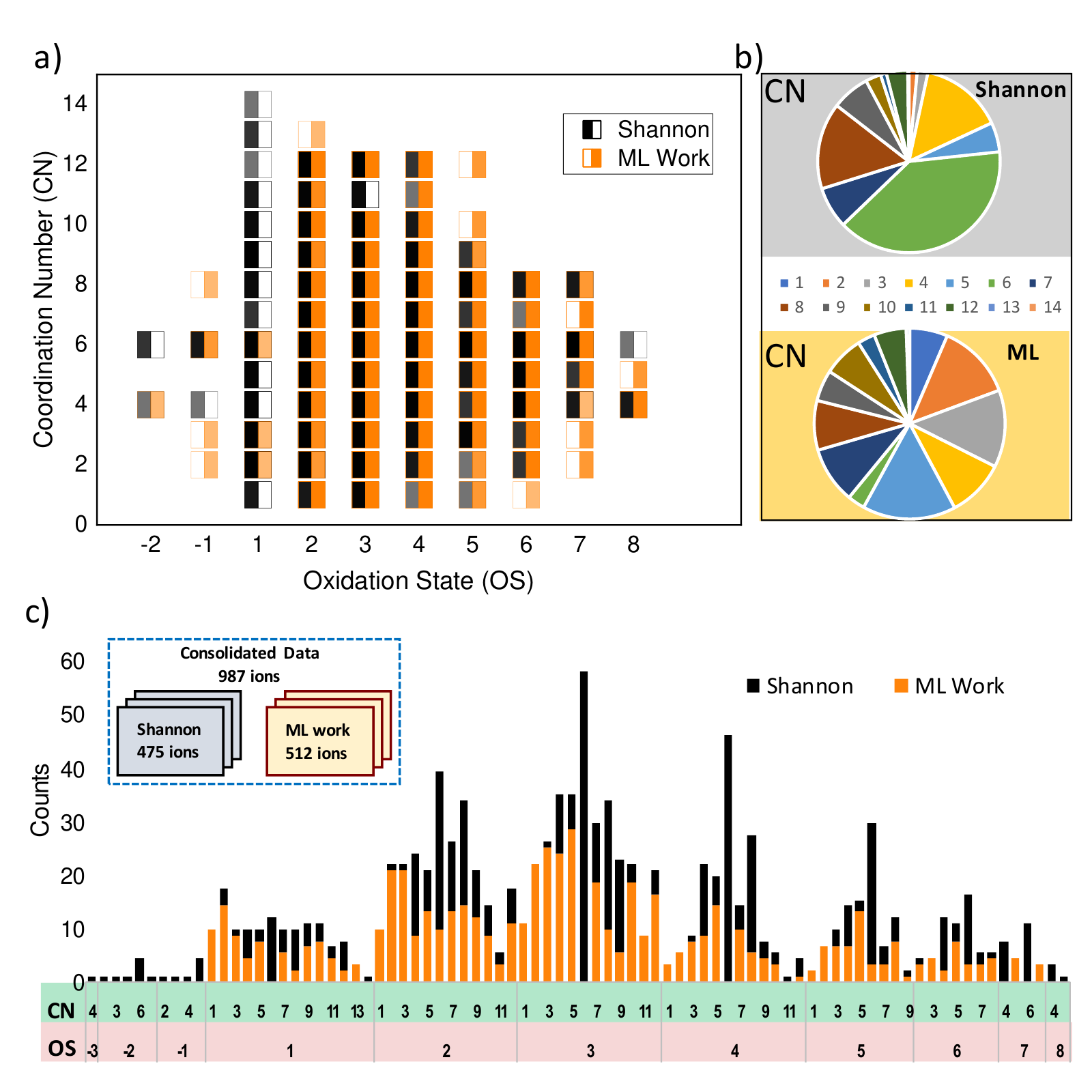}
\caption{Data comparison for the chemical environment in terms of oxidation state (OS) and coordination number (CN) from Shannon and the present ML work. a) Regions for features OS and CN covered by Shannon and ML work. b) CN: 1--14 present in both the databases considered.  c) CN present in each OS analyzed with an inset showing total Shannon and ML work data. In all these graphs, ML work consists of 512 unique ionic information harvested.}
\label{Fig2}
\end{figure*}

\subsection{Regression Procedures} State-of-the-art supervised learning models were used to develop the ionic radii ML model for Shannon’s data. The model was trained with an objective function of RMSE minimization. First, we performed 7-fold cross-validation to learn the hyperparameters and avoid overfitting. In the second stage, the best hyperparameter model for which $k$-fold reports the lowest error was then selected to test the model for prediction. Using this method, we chose the ML model with the best average prediction error. Hyperparameters in fitting the model are automatically determined internally for each regression algorithm using MATLAB \cite{R47}. The following machine learning models and their sub-classes applied for this work are mentioned below:
\begin{itemize}
    \item Linear Regression (LR): Linear, Interactions Linear, Robust Linear, and Stepwise Linear.
    \item Support Vector Machines (SVM): Linear, Quadratic, Cubic, Fine Gaussian, Medium Gaussian, and Coarse Gaussian.
    \item Decision Trees (DT): Fine, Medium, and Coarse.
    \item Ensemble of Decision Trees: Boosted Trees and Bagged Trees.
    \item Gaussian Process Regression (GPR): Squared Exponential, Exponential, Rational Quadratic, Matern 5/2 and Matern 3/2 Kernel.
\end{itemize}

\subsection{Extension methodology for ionic radius}
We analyzed the chemical information (OS and CN) for a total of 7969 crystal structures from available databases to extend the ionic radii database, as highlighted in Fig. \ref{Fig1}b. Ionic information for 7969 oxides was extracted from the Materials Project Database \cite{R30} for ICSD \cite{R27}. They are summarized (excluding duplicates) in Appendix Table-S1.  This resource provided oxidation states and coordination environments for species, which were necessary for the correct prediction of missing ions. A total of 512 unique ions (in terms of CN and OS) were found after removing the duplicates from the Shannon’s database. Accordingly, these unique ionic features consisting of new oxidation state and coordination environment along with their elemental properties were then curated for the prediction of $R_i$.

GPR model was selected based on the lowest RMSE and highest $R^2$ achieved among other regression models. The technical details of the GPR model are provided in the supplementary material. This ML model was then supplied with missing ionic properties (OS and CN) shown in Fig. \ref{Fig1}b along with their respective elemental properties vector (period, $E_{IP}$, outer shell valence electrons in – $s$, $p$, $d$, $f$ orbitals) to extend the ionic radii database. The predicted values were then consolidated with Shannon’s original data to build an up-to-date comprehensive table of 987 species and their respective ionic radii. It should be noted that in the case of an overlap, we kept the original Shannon’s empirical values, i.e., no value of Shannon is altered in the proposed improved table. Fig. \ref{Fig1}c displays the web-interface (The database of ionic radii is made open through \url{https://cmd-ml.github.io/}) in a periodic table style that was created for dissemination of the results, which can be valuable to many natural sciences.

\section{Results \& Discussion}
Prediction of the ionic radii for missing ions in Shannon’s database was performed using supervised machine learning and data harvesting from materials project. The developed GPR regression model was primarily used to predict ionic radii of rare oxidation states and coordination numbers not considered in the current ionic radii databases. Regression algorithms as function, $g(x)$, physical features of the period number, OS, CN, $\exp(-E_{IP})$ and outer shell valence electrons in $s$, $p$, $d$, $f$ orbitals were employed to Shannon’s ionic radii database in the form of 
\begin{equation}
    \label{Eq01}
    R_i=g(\textnormal{OS, CN, Period, }s,\, p,\, d,\, f,\, \exp(-E_{IP}))
\end{equation}
It is important to highlight that the ionic radii predicted from this model are extensions of Shannon’s ionic radii. 

\subsection{Data Analysis}
We have analyzed the ionic radii data for the common and uncommon oxidation states and coordination numbers by comparing Shannon’s data to current online material databases. Rare oxidation states and coordination environments, as well as those missing in Shannon’s database, are considered for extending the ionic radii database. To highlight the gaps in Shannon’s table and this study’s contribution, called “ML work” hereon, a visualization for OS and CN parameter space is provided in Fig. \ref{Fig2}a. The scatter plot shows the regions of OS and CN by Shannon and ML work with the color showing the density of the occurrences. 

Naturally, the majority of the data points in Shannon’s region covers common OS and CN whereas ML work was able to identify uncommon unreported regions of CN as shown in the pie charts of Fig. \ref{Fig2}b. Shannon’s data (total 475 ions considered) comprises primarily the following common CN geometries: CN=6 with octahedral and trigonal prism has 187 occurrences, CN=4 with tetrahedral and square planar geometry has 70 ions whereas CN=8 has a frequency of 73. These three covers 69.5\% of the total CN space reported by Shannon \cite{R16}. On the other hand, data harvested for ML work and missing ions showed primarily rare CN and OS. For instance, out of 512 new unique ions, 15.8\% were found in CN=5 with coordination geometry of trigonal bipyramidal and square pyramidal. Shannon’s data, on the contrary, had only 5.2\% of CN=5. Similarly, CN=2 for ML work had a 12.8\% occurrence whereas Shannon’s table had 1.47\% respectively in their corresponding datasets. In terms of formal charge on the ion, i.e. oxidation state, most of the data for Shannon’s table is cation as the database itself was developed using O2- anion with a six-fold coordination number and the ionic radius of 1.40 \AA. Fig. \ref{Fig2}c shows the summary for a CN connecting to a particular OS covered by the present work and Shannon. Both these sources provide a total of 987 species, with the majority of the ions found in OS=2 (23.7\%) and OS=3 (30.1\%) for both sets together. A total of 512 unique ions in terms of CN and OS were found after removing the duplicate information in the data sets as shown in Fig. \ref{Fig2}c. These new unique ionic features consisting of OS, CN and elemental properties were then employed to extend the ionic radii database using GPR.  

\subsection{Regression analysis for Shannon’s database}
Regression on the ionic radius as a function of the period number, oxidation state, coordination environment, electron affinity, ionization potential, and orbital configuration was performed for Shannon’s data. For evaluating the model accuracy and features, these measures were used:
\begin{description}
    \item[Root Mean Square Error (RMSE)]
    \begin{equation}
        \textnormal{RMSE}=\sqrt{\frac{1}{n} \sum_{i=1}^n \left( R_i - \tilde{R}_i \right)^2}
    \end{equation}
    \item[R-square] 
    \begin{equation}
        R^2=1-\frac{\sum_{i=1}^n \left( R_i - \tilde{R}_i \right)^2}{\sum_{i=1}^n \left( R_i - \bar{R}_i \right)^2}
    \end{equation}
    \item[Spearman’s rank correlation coefficient]
    \begin{equation}
        \rho=1-\frac{6 \sum_{i=1}^n d_i^2}{n^3-n}
    \end{equation}
\end{description}
Here, $n$ is the number of observations, $\tilde{R}_i$ is the predicted Shannon’s ionic radius, $\bar{R}_i$ is the mean of Shannon’s ionic radii. $d_i=\textnormal{rank}(x_i)-\textnormal{rank}(y_i)$, is the difference between the two ranks of each observation in different variables. 

\begin{figure*} [htp]
\includegraphics[width=0.9\textwidth]{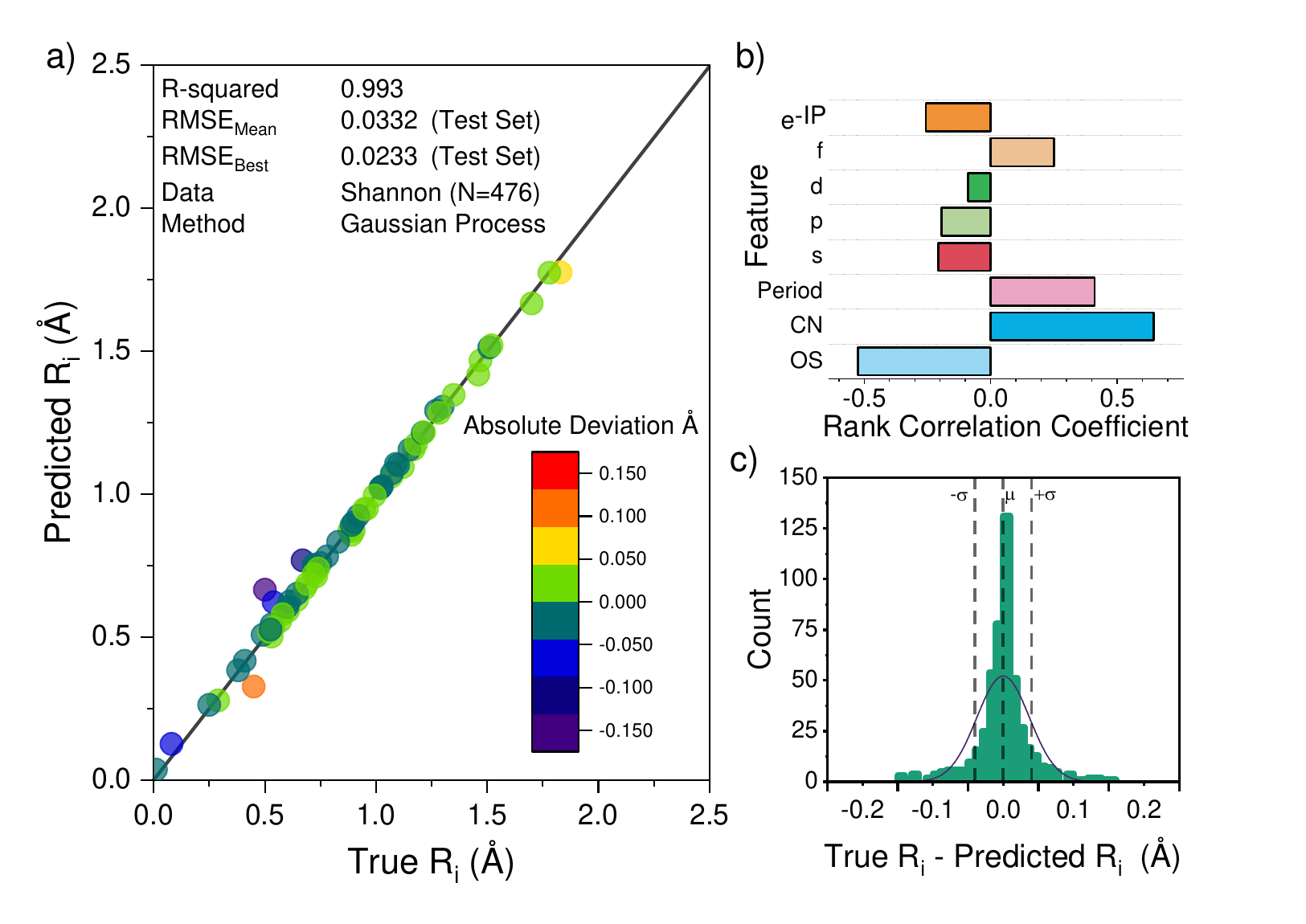}
\caption{Results from regression analysis for Shannon’s data (testing set) using a feature vector containing period number, CN, OS, outer shell valence electrons in orbitals - $s$, $p$, $d$, $f$ and $\exp(-E_{IP})$ for ionic radius prediction. a) Best performing Gaussian Process Regression (GPR) model for predicting $R_i$ with an inset showing absolute deviation error (True $R_i$– Predicted $R_i$), b) Rank correlation coefficient for $R_i$ with the feature vector and c) Distribution for absolute deviation and corresponding Standard Deviation ($\sigma$) of 0.0398 \AA.}
\label{Fig3}
\end{figure*}

Twenty different regression models were developed as highlighted in Supplementary Materials. Amongst all the developed models, the GPR model with the Matern 3/2 kernel function (details in Supplementary Materials) was the most accurate in predicting the ionic radii with an $R^2$ of 99.3\% and RMSE mean of 0.0332 \AA. Fig. \ref{Fig3}a shows the results for testing set error (7-fold validation using the fixed optimal hyperparameter model. Using the selected features, Fig. \ref{Fig3}a shows the regression results of the GPR model where it was able to achieve the minimum RMSE mean of 0.0332 \AA $\,$ over 7-folds. The results are promising as this is a first general-purpose model for all periodic elements whereas previous attempts have separately dealt with transition metals, nitrides, sulfides, etc. \cite{R34,R36,R38}. Moreover, the main advantage of GPR is that it directly captures the model uncertainty. To assess the robustness of the model, we performed 7-fold cross-validation on Shannon’s data of 475 ions where the hyperparameters of the model were optimized using the quasi-newton approach with a function tolerance of $10^{-6}$ for Matern 3/2 kernel. The color bar in Fig. \ref{Fig3}a shows the absolute deviation $|R_i-\tilde{R}_i|$ for each point where the majority of the data points lie in the standard deviation of +/- 0.0398 \AA as shown in the inset figure. To validate the significance of the eight features -- period number, OS, CN, $\exp(-E_{IP})$, and orbital configuration $s$, $p$, $d$, $f$ -- we assess the Spearman’s rank correlation coefficient ($\rho$) between individual feature and ionic radii as shown in Fig. \ref{Fig3}b. In the rank correlation, the maximum correlation happens at a value of 1 whereas the direction is shown by the positive or negative sign. Interestingly, the maximum correlation, $\rho=+0.64$, was found for CN. It shows that with an increase in CN, $R_i$ also increases. OS was found to have $\rho=-0.52$ showing a negative correlation with $R_i$. This is because as the OS increases, atoms lose electrons hence the overall effective nuclear charge increases resulting in reduced $R_i$. Period number resulted in a value of +0.41 due to the addition of outermost shell which causes the $R_i$ to increase. This was followed by $\exp(-E_{IP})$ with $\rho=-0.25$. Fig. \ref{Fig3}c shows that the absolute deviation from the model follows normal distribution, signifying that there is no systematic error present in the developed regression method.

\begin{figure*} [htp]
\includegraphics[width=0.9\textwidth]{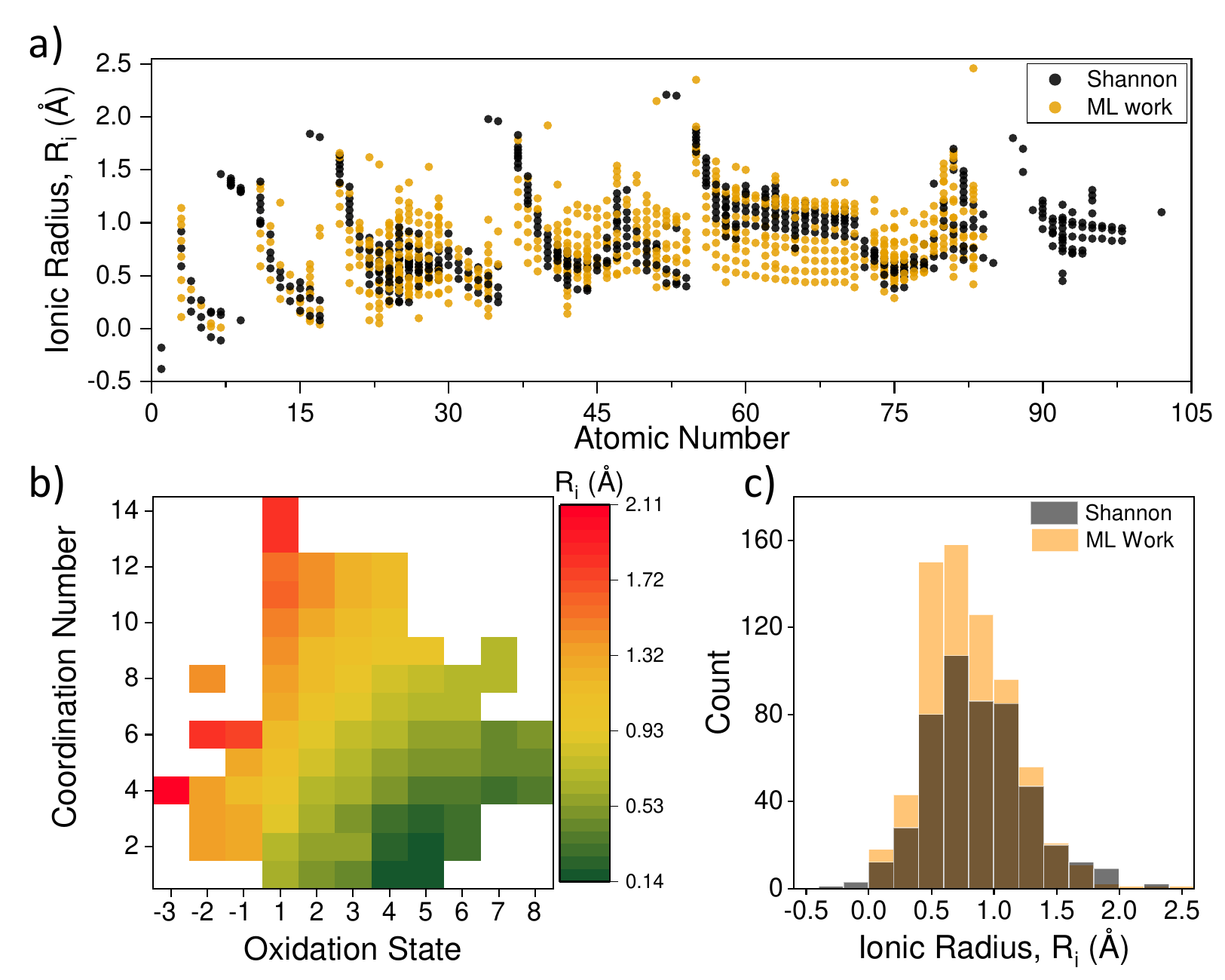}
\caption{Extension to new ions with results from prediction and consolidated database, a) Ionic radii predicted from ML work (GPR model) merged with Shannon’s database as a function of atomic number, b) General behavior of ionic radius as a function of coordination number and oxidation state for N=987 consolidated species, c) Histogram of Shannon’s empirical database and current ML work showing the similarity of the ionic radii distribution.}
\label{Fig4}
\end{figure*}

\subsection{Extension to missing ions}
We performed an extension of the new ions using the GPR prediction model on a set of 512 ions as depicted in Fig. \ref{Fig4}a. The completed consolidated table is shown in Appendix Table-S2 with the feature vector and ionic radii values. Here, we kept the original Shannon’s empirical values where we found an overlap, i.e., no ionic radius from Shannon was changed in the proposed table. Fig. \ref{Fig4}a shows that the predicted elements were inclusive in the range of atomic numbers by Shannon (from 1 to 102). In Fig. \ref{Fig4}b, we see the effect of the oxidation state and coordination number on the total consolidated dataset. The radii decrease with an increase in the oxidation state due to the additional effective nuclear charge by losing an electron. The ionic radius was found to increase with higher CN because the electron field is stretched out by the existence of additional surrounding ions for higher dimensional polygons.

Fig. \ref{Fig4}c shows the distribution of ionic radius predicted and Shannon’s original database. The shape of these histograms shows that the predicted $R_i$ is comparable to the dataset used for Shannon’s ionic radii. The difference is primarily due to the number of counts and rare OS/CN in our feature vectors. The descriptors are relatively simple and physically intuitive, having only features highlighting its robustness for extending to new ions as they appear in online databases. The completed consolidated table is uploaded at the open materials database website. A comparison of ionic radii for cations calculated using Brown and Shannon (BS) \cite{R48},Shannon \cite{R42}, Ouyang \cite{R40} and current ML work is presented in Fig. S2. It should be noted that as the code can predict ionic radii for any arbitrary OS/CN, care must be taken when selecting chemical environment information for realistic ionic radii calculation.

\subsection{Further analysis}
Our predictive model was extended to new ions based on compounds that were primarily experimentally observed. According to our statistical analysis, about 96\% of these compounds (7658 of 7969) are experimentally associated with multiple Inorganic Crystal Structure Database (ICSD) IDs \cite{R49} (the complete data is provided in Supplementary Table-S4). Hereunder, arbitrarily selected ions from our predicted ionic radii dataset are discussed to illustrate the importance of our predictive model. For crystal structure classification, revised ionic radii have been adopted for accurate tolerance factor predictions \cite{R20}. Searching for \ch{Sm^2^+} based perovskites, Travis et. al. \cite{R20} had to apply \ch{Sm^2^+} (CN-7) ionic radius as Shannon’s table provides no 6 coordinate \ch{Sm^2^+} ion[20]. An experimental investigation of \ch{BiCoO3} with pyramidal polar coordination of \ch{Co^3^+} (CN-5) under high pressure was implemented and spin transition was observed due to the change of \ch{Co^3^+} (CN-5) in the atmospheric pressure phase to the approximately isotropic octahedral coordination (CN-6) in the high-pressure phase \cite{R50}; this is not covered in the original Shannon table. Also, the structural and magnetic properties of \ch{LiRO2} (R= rare-earth) were experimentally studied by Hashimoto et. al. \cite{R51}. The X-ray diffraction measurements have shown that the \ch{LiErO2} compound was found to form $\beta$-type (space group: P21/c) with \ch{Li^+} (CN-3) below the room temperature \cite{R51}; this is not covered in the original Shannon table as well. For octahedral coordination of \ch{Sn^2^+} with (CN-6), there are efforts to stabilize the \ch{CsSnCl3} perovskite structure, which prefers to form pyramidal coordination instead of octahedral coordination, by ionic substitution of \ch{Sn^2^+} with smaller octahedral cations \cite{R52}. The octahedral coordination \ch{Sn^2^+} was used as a possible candidate that replace \ch{Pb^2^+} in perovskite structure due to its toxicity \cite{R52}. This experimentally observed OS of \ch{Sn} is not included in the original Shannon’s table. Table \ref{T1} lists the considered oxidation states that are not tabulated in Shannon’s table. For a concise reference, all of them are among the listed oxidation state in the seminal book by Greenwood and Earnshaw \cite{R53}. Many other references can be found for each one of them. Nonetheless, they are certainly not among the common oxidation states but they cannot be ignored.

\begin{table}
    \centering
    \begin{tabular}{|p{0.12\textwidth}|p{0.12\textwidth}|p{0.12\textwidth}|}
    \hline
        C: +2, +3 & P: +4 & S: +2, +3, +5 \\
    \hline
        Cl: +1, +4 & Sc: +2 & Fe: +5 \\
    \hline
        Co: +1 & Ni: +1 & Ge: +3 \\
    \hline
        Se: +2 & Y: +2 & Nb: +2 \\
    \hline
        Ru: +6 & In: +1, +2 & Sn: +2 \\
    \hline
        Te: +5 & La: +2 & Ce: +2 \\
    \hline
        Pr: +2 & Gd: +2 & Ir: +6 \\
    \hline
        Pt: +6 & Au: +2 & Th: +3 \\
    \hline
    \end{tabular}
    \caption{The considered oxidation states that are not tabulated in Shannon’s table}
    \label{T1}
\end{table}

\section{Conclusion}
A very rigorous and highly accurate Machine Learning approach is employed to extend the renowned Shannon’s table from 475 ions to 987 ions. In ML implementation, the original Shannon's table is used to develop the ionic-radius regression model as a function of the period number, the valence orbital configuration, OS, CN, and Ionization Potential. The model is then implemented to extend the ionic radii table for all possible combinations of OS and CN available in crystallographic repositories. Many ML methods are considered and a comparison was carried out. 
In the Gaussian Process Regression (GPR) model, the reached $R^2$ accuracy is 99\% while the root mean square error of radii is 0.0332 \AA. The generated data is consolidated with the reputable Shannon’s data and is made available online in a database repository \url{https://cmd-ml.github.io/}.

\begin{acknowledgments}
SM Alqahtani would like to thank he Ministry of Education in the Kingdom of Saudi Arabia
for the financial support by its Research and
Development Office (RDO).
\end{acknowledgments}

\bibliography{IRbib}

\end{document}